# The Age of Updates in a Simple Relay Network

Ali Maatouk*, Mohamad Assaad*, and Anthony Ephremides†

*TCL Chair on 5G, Laboratoire des Signaux et Systèmes, CentraleSupélec, Gif-sur-Yvette, France
†ECE Dept., University of Maryland, College Park, MD 20742

*Abstract*—In this paper, we examine a system where status updates are generated by a source and are forwarded in a First-Come-First-Served (FCFS) manner to the monitor. We consider the case where the server has other tasks to fulfill, a simple example being relaying the packets of another stream. Due to the server's necessity to go on vacations, the age process of the stream of interest becomes complicated to evaluate. By leveraging specific queuing theory tools, we provide a closed form of the average age for both streams which enables us to optimize the generation rate of packets belonging to each stream to achieve the minimum possible average age. The tools used can be further adopted to provide insights on more general multi-hop scenarios. Numerical results are provided to corroborate the theoretical findings and highlight the interaction between the two streams.

## I. INTRODUCTION

The Age of Information (AoI) is a new concept that has been introduced in [1] and is considered of broad interest in communication systems. More particularly, ubiquitous connectivity and the cheap hardware cost have created new applications where sensors are able to send status updates to a certain receiver. The status can range from being as simple as the temperature of a room [2] to as complicated as a vehicle's position and velocity [3]. In these applications, a source generates time-stamped updates that are forwarded through a network to reach the intended monitor. These updates should therefore be as timely as possible, i.e., the receiver should have the freshest possible information about what is being monitored. Knowing that the transmission of the packets may include channel errors and back-off timers to mitigate interference from neighboring sensors, the characterization of this metric is far from being straightforward.

Although practical scenarios of interest can be quite complex, the investigation of this metric for even the simplest models was found to be challenging. In [1], the average age was formulated for the case of First-Come-First-Served (FCFS) disciplines: M/M/1, M/D/1 and D/M/1 where it was shown that an optimal packets generation rate can be found to minimize the average age. The case where multiple information sources share the same queue was also evaluated in [4].

In [5], it was shown that if the source had the ability to manage packets, the average age of the stream can be further reduced. Moreover, a related metric called the *average peak age* was introduced in [5]. With this metric being generally more tractable than the average age, the authors in [6] formulated the average peak age of information for multiple information sources under general service time distributions.

With the AoI metric gaining more attention, a surge of papers have been recently published on the subject [7]. The age of information for energy harvesting sources was intesively investigated in the literature (e.g. [8]). More recently, a shift of interest to multi-hop scenarios can be witnessed. This can be justified by the fact that the AoI is of high interest in wireless sensor networks where, in general, sensors can be multiple hops away from the monitor. In [9], it was shown that when the transmission times of packets in the network are exponentially distributed across all nodes, the Last-Come-First-Served (LCFS) preemptive policy at the relaying nodes minimizes the average age of the stream. Although this work gives insights on the disciplines to be used, it does not provide ways to explicitly calculate the average age of each stream. More recently, an explicit calculation of the average age was calculated for a single information stream over a network of preemptive servers in [10]. In another work [11], due to the complexity of the average age calculations, an upper bound and lower bound were provided for the case where a server has two streams to serve: a FCFS stream and a LCFS with preemption stream.

In the present work, we overcome the complexity that was found in [11] by leveraging some key queuing theory tools. By doing so, a study of a simple relay scenario will be feasible as will be depicted in the sequel. More precisely, we are able to formulate a *closed form* of the average age of both streams along with a discussion on the interaction between the two streams in terms of both stability and their respective age. In fact, it will be shown in the sequel that the minimization of the age of the stream of interest admits a unique optimal packets generation rate. The interest in those tools used in the formulation come from the fact that they can be used to evaluate even more complicated multi-hop scenarios. Moreover, we study the case where the second stream (i.e. the relayed packets) is not age-sensitive. This comes from the fact that some applications are more restricted by throughput than age. In this case, we aim to find the best way to minimize its effect on the age sensitive stream. It is worth mentioning that our system model is general and that is it not bound to the relay scenario taken into account. It can be used for any type of scenarios where the server is allowed to be unavailable for a certain amount time. Some non-exclusive examples include the node transitioning to SLEEP mode or doing necessary offline processing.

The paper is organized as follows: Section II describes the system model. Section III presents the theoretical results on the average age of the streams. Section IV provides numerical results that corroborate the theoretical findings while Section V concludes the paper.

## II. SYSTEM MODEL

Consider an information source sending status updates to a monitor. In this scenario, the instantaneous age of information at the receiver at time instant $t$ is defined as:

$$\Delta(t) = t - U(t) \qquad (1)$$

where $U(t)$ is the time stamp of the last successfully received packet by the monitor. Suppose that packet $j$ is generated at time instant $t_j$ and is received by the monitor at time $t'_j$, the evolution of the instantaneous age in this case can be depicted in Fig. 1 where $A_j$ and $T_j$ denote the inter-arrival and sojourn time of packet $j$ respectively.

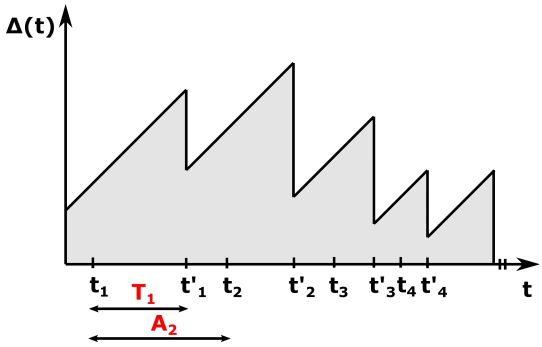

Fig. 1: The time evolution of the instantaneous age $\Delta(t)$

The main interest lays in the computation, and eventual minimization, of the average age of information in the aim of keeping the information as fresh as possible at the monitor. The average age is nothing but the saw-tooth area highlighted in Fig.1 and is defined as:

$$\Delta = \lim_{\tau \to \infty} \frac{1}{\tau} \int_0^\tau \Delta(t) dt \qquad (2)$$

The computation of this metric is however challenging, even in the simplest scenarios [1]. With the metric being relatively new, giving insights on the computation and minimization of the average age in particular realistic scenarios is of paramount importance.

The scenario of interest that we investigate in the following consists of an information source generating packets, denoted as stream 1, and buffering them in a single queue (FCFS discipline) where the packet $j$ inter-arrival time $A_j$ is supposed to be exponentially distributed with mean $\frac{1}{\lambda_1}$. In other words, $\lambda_1$ can be seen as generation rate of packets belonging to stream 1. The physical service time of the packet $G_j$ is supposed to be exponential as well with mean $\frac{1}{\mu_1}$. However, we consider the case where the server is not always available and can take eventual vacations. When a vacation takes place, the transmission of the packet of stream 1 is interrupted (with eventual resume) until the vacation finishes. The decision to go on vacation is dictated by an appropriate timer $S$ that is supposed to be exponentially distributed with mean $\frac{1}{s}$. The vacation duration $W$ is also supposed to be exponentially distributed with mean $\frac{1}{w}$. What makes this model of broad interest is the fact that, in realistic scenarios, a device may have different functionality than just being occupied by forwarding its own generated packets. Some non-exclusive examples of vacations are:

- The device may transition into $SLEEP$ mode for a certain amount of time after which it wakes up. The motivation behind this transition is to prolong the battery's life
- The device may proceed to do necessary offline processing
- The device may relay packets of different streams other than its own

In the sequel, for clearer presentation, we focus on the case where the vacation is motivated by relaying packets of different streams although the analysis is still valid for any of the aforementioned cases. Sensor 1 is considered as the sensor of interest that is generating age-sensitive data and forwarding them to a monitor that is only one hop away. However, at the same time, it keeps a separate queue for packets generated by sensor 2 in the aim of forwarding them to the monitor as seen in both Fig. 2 and 3.

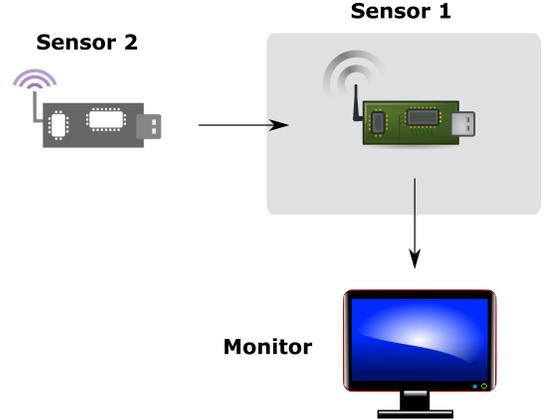

Fig. 2: A simple relay network

We consider in the following work two different applications for the simple relay network in question:

- The relayed stream 2 is not age-sensitive and we are simply restricted by a certain target average throughput
- The relayed stream 2 is age-sensitive

We first start by providing the general mathematical formula used to calculate the age of the stream of interest (stream 1). When the system is in steady state, both the inter-arrival and the sojourn times of the packets of stream 1 are stochastically identical, i.e., $A_j \stackrel{\text{st}}{=} A_{j-1} \stackrel{\text{st}}{=} A$ and $T_j \stackrel{\text{st}}{=} T_{j-1} \stackrel{\text{st}}{=} T$. We therefore lose the packet's index $j$ in the sequel. In this case, the average age of stream 1 is evaluated as [1]:

$$\Delta_1 = \lambda_1(\mathbb{E}(AT) + 1/\lambda_1^2) \qquad (3)$$

The main difficulty in finding a closed form of the average age of a certain stream lays in the term $\mathbb{E}(AT)$. In fact, the two variables are not independent (a large inter-arrival time allows the system to be emptied and $T$ to be smaller) and the evaluation of this term, even in the simplest scenarios, can be challenging. In the following, we provide a closed form expression of the average of the stream of interest. Using the formulated expression, insights on the effect of the vacation is highlighted along with finding the optimal packets generation rate $\lambda_1^*$ to minimize the average age of stream 1 is provided. To do so, we first model the system using a $2D$ Markov chain. We then proceed to leverage the notion of probability generating functions and Little's distributional law. After appropriate analysis, the closed form is formulated along with different discussions on practical scenarios considerations.

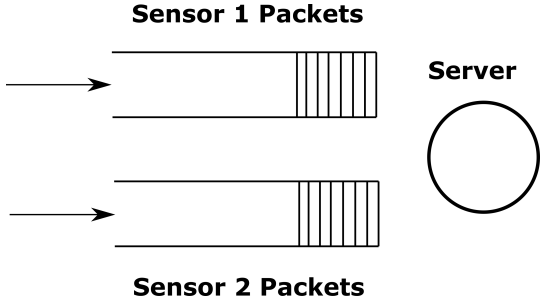

Fig. 3: The perspective of Sensor 1

## III. AGE CALCULATION

Let us consider the 2D continuous time stochastic process $\{(N(t), M(t)) : t \geq 0\}$ where $N(t) \in \mathbb{N}$ and $M(t) \in \{V, B\}$ denote the number of packets of stream 1 inside the system and the server status at time instant $t$ respectively. $M(t)$ can only take two values: 1) $V$ when the server is on vacation and 2) $B$ when the server is not on vacation and is serving or is ready to serve the stream of interest. By taking into account the model's assumptions, $\{(N(t), M(t)) : t \geq 0\}$ becomes a Markovian process and the evolution of the system can be represented by a 2D continuous time Markov chain as seen in Fig. 4 where the states are all the possible combinations of $(N, M)$.

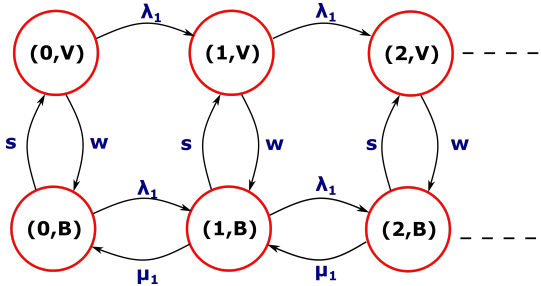

Fig. 4: Markov Chain

Let $\pi_B(i)$ be the stationary distribution for the state where the server is not on vacation while $i$ packets of stream 1 are in the system. On the counterpart, let $\pi_V(i)$ be the case where the server is on vacation with $i$ packets of stream 1 in the system. Based on the preceding, we can define the Probability Generating Function (**p.g.f**) of the random variable $N$ as:

$$P(x) = \mathbb{E}(x^N) = \sum_{i=0}^{\infty} \pi(i) x^i = \sum_{i=0}^{\infty} \pi_V(i) x^i + \sum_{i=0}^{\infty} \pi_B(i) x^i = P_V(x) + P_B(x).$$

**Theorem 1.** *In the aforementioned system, the p.g.f of the number of packets of stream 1 in the system is given by:*

$$P(x) = \frac{\mu_1 \pi_B(0)(\lambda_1 + w + s - \lambda_1 x)}{\lambda_1^2 x^2 - (\lambda_1 w + \lambda_1^2 + \lambda_1 s + \lambda_1 \mu_1)x + \lambda_1 \mu_1 + w\mu_1} \quad (4)$$

*where*[1]:

$$\pi_B(0) = \frac{\mu_1 - \lambda_1 - \lambda_1 \frac{s}{w}}{\mu_1(1 + \frac{s}{w})} \quad (5)$$

*Proof:* The proof can be found in Appendix A. ∎

**Theorem 2.** *The sojourn time $T$ of packets belonging to stream 1 in the system has the following distribution:*

$$f_T(t) = C_1 \exp(\alpha_1 t) + C_2 \exp(\alpha_2 t) \quad (6)$$

*where $\alpha_1$ and $\alpha_2$ are the roots of the second degree equation:*

$$z^2 + z(w + s + \mu_1 - \lambda_1) + w\mu - \lambda_1 w - \lambda_1 s = 0 \quad (7)$$

*Moreover, $C_1$ and $C_2$ are given by:*

$$C_1 = \frac{(\mu_1 - \lambda_1 - \lambda_1 \frac{s}{w})(w + s + \alpha_1)}{(1 + \frac{s}{w})(\alpha_1 - \alpha_2)} \quad (8)$$

$$C_2 = \frac{(\mu_1 - \lambda_1 - \lambda_1 \frac{s}{w})(w + s + \alpha_2)}{(1 + \frac{s}{w})(\alpha_2 - \alpha_1)} \quad (9)$$

*Proof:* The proof can be found in Appendix B. ∎

### A. Application 1: Relay of non age sensitive data

Suppose that the packets that are being relayed are not age sensitive and our aim is to have a desired average throughput for stream 2. The average throughput experienced by the relayed stream is equal to the portion of time the server spends on vacation[2].

**Corollary 1.** *The average amount of time the server spends on vacation is:*

$$\pi_V = \frac{\frac{s}{w}}{\frac{s}{w} + 1} \quad (10)$$

*Proof:* The proof can be found in Appendix C. ∎

We can see from the results of this corollary that the average throughput of stream 2 depends on the ratio of $\frac{s}{w}$ and not individually on them. These results are of paramount interest: In fact, investigating the average age of stream 1, for a *fixed* $\frac{s}{w}$, answers the important question; which has the least effect on the age of stream 1, short and numerous vacations, or long rare vacations? If it turns out to be the first, then packets of stream 2 are better off being served in fragments by assigning short mean vacation times. If it is the latter, then packets of stream 2 are better served as a *batch* (i.e assigning long mean vacation

---

[1]The system is stable if and only if $\pi_B(0) > 0$ which translates into having $\mu_1 > \lambda_1 + \lambda_1 \frac{s}{w}$

[2]It is assumed that the queue corresponding to stream 2 is always backlogged, i.e., whenever the server goes on vacation, it finds a packet to relay. If not, dummy packets are supposed to be sent which is a natural assumption to evaluate the average throughput performance [12][13][14, p.8]

time which would result in numerous packets being served in a single vacation). In both cases, the average throughput of stream 2 is the same as the frequency of the vacations is different. To be able to answer this question, we provide a closed form of the average age of stream 1 in the following.

**Proposition 1.** *The average age of the stream of interest is given by:*

$$\Delta_1(\lambda_1, \mu_1, s, w) = \frac{\lambda_1^2 C_1}{\alpha_1^2(\alpha_1 - \lambda_1)^2} + \frac{\lambda_1^2 C_2}{\alpha_2^2(\alpha_2 - \lambda_1)^2} + \frac{1}{\lambda_1}$$
$$+ \frac{\mu_1 s(\mu_1 - \lambda_1 - \lambda_1 \frac{s}{w}) + (s+w)(\lambda_1 + w)(\mu_1(1 + \frac{s}{w}))}{(w\mu_1)(\lambda_1 + w)(\mu_1(1 + \frac{s}{w}))} \quad (11)$$

*Proof:* The proof can be found in Appendix D. ∎

The next step revolves around finding, for a fixed setting $(\mu_1, s, w)$, the optimal packets generation rate $\lambda_1^*$ to minimize the average age of stream 1. Insights on this optimization is provided in the following proposition.

**Proposition 2.** *For a fixed service rate $\mu_1 > 0$, mean vacation time $\frac{1}{s} > 0$ and mean vacation duration $\frac{1}{w} > 0$, the following optimization problem admits a unique minimizer $\lambda_1^*$:*

$$\begin{aligned} \underset{\lambda_1}{\text{minimize}} \quad & \Delta_1(\lambda_1) \\ \text{subject to} \quad & 0 < \lambda_1 < \frac{\mu_1}{1 + \frac{s}{w}} \end{aligned} \quad (12)$$

*Proof:* The proof can be found in Appendix E. ∎

By using the results of Proposition 2, we can answer the question previously stated as for the nature of vacations that have the least effect on the average age of stream 1. As we have shown in the proof in Appendix E, the optimal solution of the problem (12) in the interval $]0, \frac{\mu_1}{1+\frac{s}{w}}[$ can be achieved by finding $\lambda_1^*$ such that $\frac{d\Delta_1(\lambda_1)}{d\lambda_1}|_{\lambda_1=\lambda_1^*} = 0$ (please refer to Appendix E for more details). After formulation of $\lambda_1^*$, we can compare the minimal achievable average age $\Delta_1(\lambda_1^*, \mu_1, s_1, w_1)$ and $\Delta_1(\lambda_1'^*, \mu_1, s_2, w_2)$ such that $s_1 > s_2$ and $w_1 > w_2$ while preserving $\frac{s_1}{w_1} = \frac{s_2}{w_2}$ to find the answer to our question. It turns out, as will be seen in the numerical results section, that it is better to have numerous short vacations rather than long rare vacations to reduce the effect of the vacations on the average age of stream 1.

### B. Application 2: Relay of age sensitive packets

As previously mentioned in the introduction, the average age of information in multi-hop networks was investigated in [9] and it was shown that, for exponential packets transmission time, relayed packets should follow LCFS policy with preemption to minimize their age. In this case, the vacation can be thought to be an arrival of a packet of stream 2 that preempts any packet being transmitted, either from stream 1 or stream 2. In this case, the relayed packets see the system as an M/M/1/1 queue and their average age is therefore [4]:

$$\Delta_2 = \frac{1}{\mu_2} + \frac{1}{\lambda_2} \quad (13)$$

where $\lambda_2$ and $\mu_2$ are the arrival and service rate of stream 2 respectively. This type of scenarios was investigated in [11] where a lower and upper bound on the average age of stream 1 was provided using the detour flow graph tool. Unlike the work in [11], we develop a *closed form* of the average age of stream 1 and give insights on choosing the optimum packets generation rate $\lambda_1$ to minimize the age of stream 1. This highlights the importance of the tools presented in the Appendices A and B as they are able to reduce the difficulty of the average age calculations for certain complex scenarios.

**Proposition 3.** *The average age of stream 1 is given by replacing $s = \lambda_2$ and $w = \mu_2$ in eq. (11)*

$$\Delta_1(\lambda_1, \mu_1, \lambda_2, \mu_2) = \frac{\lambda_1^2 C_1}{\alpha_1^2(\alpha_1 - \lambda_1)^2} + \frac{\lambda_1^2 C_2}{\alpha_2^2(\alpha_2 - \lambda_1)^2} + \frac{1}{\lambda_1}$$
$$+ \frac{\mu_1 \lambda_2 (\mu_1 - \lambda_1 - \lambda_1 \frac{\lambda_2}{\mu_2}) + (\mu_2 + \lambda_2)(\lambda_1 + \mu_2)(\mu_1(1 + \frac{\lambda_2}{\mu_2}))}{(\mu_2 \mu_1)(\lambda_1 + \mu_2)(\mu_1(1 + \frac{\lambda_2}{\mu_2}))}$$
$$(14)$$

*Proof:* The analysis is similar to that of Appendix D. In this scenario, the vacation starts when a packet from stream 2 arrives to the system and preempts the service of packets of stream 1. The vacation finishes once the packet of stream 2 that is being served finishes. The same Markov chain of Fig. 4 holds, by replacing $s = \lambda_2$ and $w = \mu_2$. However, the difference between the two scenarios is the fact that new packets of stream 2 can preempt the previous stream 2 packet in service. In other words, the vacation duration timer can be reset due to a new arrival from stream 2. Due to the memoryless property of the exponential vacation timer, this reset of the vacation timer does not affect stream 1 and the same sojourn and virtual service time for packets of stream 1 as in Application 1 is to anticipated and the same proof of Appendix D holds. ∎

## IV. NUMERICAL RESULTS

This section provides numerical results to give insights on the theoretical conclusions previously stated.

### A. Application 1

In this scenario, we have a fixed average service rate of stream 1 and is set to $\mu_1 = 1$. Moreover, we have a certain average throughput requirement for the relayed stream to ensure its queuing stability. This requirement is supposed to be equal to $\frac{1}{2}$ in the sequel, i.e., the portion of time where the server should be relaying packets from stream 2 is equal to $\frac{1}{2}$. One question arises: knowing that the throughput depends on the ratio of $\frac{s}{w}$ rather than individually on them, how could we calibrate these two parameters in a way to have the least impact on the average age of stream 1?

For this purpose, we simulate different scenarios of $s$ and $w$ such that we keep achieving the same required average throughput ($\frac{s}{w} = 1$). As seen in Fig. 5, for the same ratio $\frac{s}{w} = 1$, it is better to opt for high values of $s$ and $w$. In other words, as previously mentioned, it is better to go often on vacation but for short amount of time. Consequently, packets of stream 2 are better served as fragments by making $w$ be as high as practically possible while providing a vacation rate $s$ high enough to keep the same ratio of $\frac{s}{w} = 1$. Moreover, there is a certain optimal rate $\lambda_1^*$ that achieves the lowest possible average age of stream 1 for a fixed setting. Discussions on this matter will be provided in the next subsection.

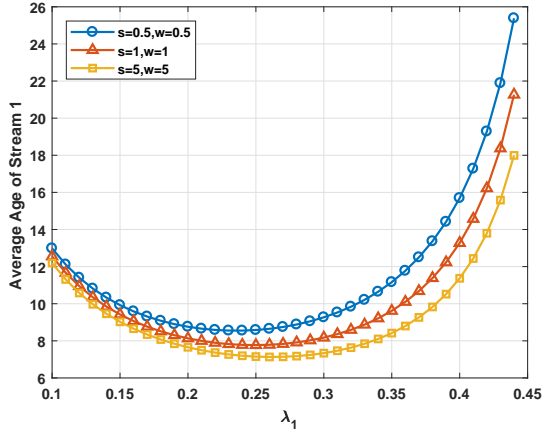

Fig. 5: Illustration of the effect of $s$ and $w$ on the average age of stream 1

### B. Application 2

In this scenario, we have a fixed average service rate of $\mu_1 = 1$ and fixed vacation timer $w = \mu_2 = 4$ which is dictated by the size of stream 2 packets. In this case $\Delta_2 = \frac{1}{4} + \frac{1}{\lambda_2}$, therefore once can clearly that increasing $s = \lambda_2$ reduces the average age of stream 2. Our goal becomes to find the packets generation rate that minimizes the age of stream 1 for a certain fixed relayed packets rate $\lambda_2$. This can be thought as an optimal packets generation rate $\lambda_1^*$ that mitigate the effect of the relay of stream 2 as much as possible. In fact, one can see that as the arrival rate $\lambda_2$ of stream 2 grows larger, the optimum $\lambda_1^*$ grows smaller and the achieved minimal average age is larger. This is due to the server being unavailable for a higher duration of time and therefore, delays are anticipated by waiting for vacations to finish which results in a necessity to reduce the packets generation rate $\lambda_1$ to mitigate this delay and therefore reduce the average age.

## V. CONCLUSION

In this paper, we have investigated the age of information in the case where the server is not always available and is allowed to go on vacation. This setting is of broad interest as it incorporates several practical scenarios, such as the relay of packets from different streams. By leveraging several key queuing theory concepts, we were able to provide a closed form expression of the average age of the stream of interest. Insights on the theoretical results were provided to further understand the interaction between the age of the stream of interest and the vacations considered. The authors believe that the analysis provided in the paper can be used to explore more general scenarios where more than one stream are relayed by the node of interest and therefore be a basis for multi-hop systems analysis.

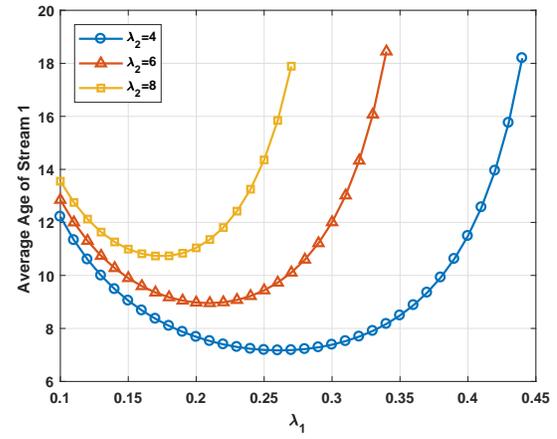

Fig. 6: Illustration of the optimal packets generation rate

## APPENDIX A
### PROOF OF THEOREM 1

We start the proof by first providing the balance equations of the Markov chain at the states characterized by $M = V$:

$$(\lambda_1 + w)\pi_V(i) = \lambda_1 \mathbb{1}\{i \neq 0\}\pi_V(i-1) + s\pi_B(i) \quad \forall i \in \mathbb{N} \tag{15}$$

where $\mathbb{1}\{i \neq 0\} = 1$ if and only if $i \neq 0$. We proceed by multiplying eq. (15) by $x^i$ and summing over all the possible values of $i \in \mathbb{N}$:

$$(\lambda_1 + w)\sum_{i=0}^{+\infty} x^i \pi_V(i) = \sum_{i=0}^{+\infty} \lambda_1 \mathbb{1}\{i \neq 0\} x^i \pi_V(i-1) + s\sum_{i=0}^{+\infty} x^i \pi_B(i) \tag{16}$$

By manipulating the power series on the right handside (change of variables, appropriate multiplication/division by the variable $x$), and by making use of the definition of the p.g.f, we can conclude the following:

$$P_V(x) = \frac{s}{\lambda_1 + w - \lambda_1 x} P_B(x) \tag{17}$$

The next step consists of formulating the balance equations for the states where the server is not on vacation:

$$(\lambda_1 + s + \mu_1 \mathbb{1}\{i \neq 0\})\pi_B(i) = \mu_1 \pi_B(i+1) + \lambda_1 \mathbb{1}\{i \neq 0\}\pi_B(i-1) + w\pi_V(i) \tag{18}$$

As it has been previously done, we multiply by $x^i$ and sum over all the possible values of $i \in \mathbb{N}$:

$$(\lambda_1 + s + \mu_1)\sum_{i=0}^{+\infty} x^i \pi_B(i) - \mu_1 \pi_B(0) = \mu_1 \sum_{i=0}^{+\infty} x^i \pi_B(i+1)$$
$$+ \lambda_1 \sum_{i=0}^{+\infty} x^i \mathbb{1}\{i \neq 0\}\pi_B(i-1) + w\sum_{i=0}^{+\infty} x^i \pi_V(i) \tag{19}$$

After some manipulations, we get the following:

$$(\lambda_1 + s + \mu_1 - \lambda_1 x - \frac{\mu_1}{x})P_B(x) = \mu_1 \pi_B(0) - \frac{\mu_1}{x}\pi_B(0) + wP_V(x)$$

By using the results of eq.(17) and the fact that $P(x) = P_V(x) + P_B(x)$, the p.g.f of the number of packets of stream 1 in the system, referred to as $N$, is:

$$P(x) = \frac{(x\mu_1 \pi_B(0) - \mu_1 \pi_B(0))(\lambda_1 + w - \lambda_1 x + s)}{(\lambda_1 + w - \lambda_1 x)(\lambda_1 x + sx + \mu_1 x - \lambda_1 x^2 - \mu_1) - wsx} \tag{20}$$

What remains is to find the constant $\pi_B(0)$. To do so, we eliminate the common factor $(x-1)$ from the numerator and denominator by taking into account that:

$$(\lambda_1 + w - \lambda_1 x)(\lambda_1 x + sx + \mu_1 x - \lambda_1 x^2 - \mu_1) - wsx = (x-1)$$
$$(\lambda_1^2 x^2 - (\lambda_1 w + \lambda_1^2 + \lambda_1 s + \mu_1 \lambda_1)x + (\lambda_1 \mu_1 + w\mu_1)) \tag{21}$$

By doing so, the p.g.f can be rewritten as follows:

$$P(x) = \frac{\mu_1 \pi_B(0)(\lambda_1 + w + s - \lambda_1 x)}{\lambda_1^2 x^2 - (\lambda_1 w + \lambda_1^2 + \lambda_1 s + \lambda_1 \mu_1)x + \lambda_1 \mu_1 + w\mu_1} \tag{22}$$

Knowing that $P(1) = \sum_{i=0}^{\infty} \pi_V(i) + \sum_{i=0}^{\infty} \pi_B(i) = 1$, we can conclude by replacing $x$ by the value of 1 in eq. (22) that:

$$\pi_B(0) = \frac{\mu_1 - \lambda_1 - \lambda_1 \frac{s}{w}}{\mu_1(1 + \frac{s}{w})} \tag{23}$$

which concludes our proof.

## APPENDIX B
### PROOF OF THEOREM 2

In this proof, we leverage the concept of Little's distributional law [15]. The theorem provides a relationship between the p.g.f of the number of packets in an ergodic queueing system and the sojourn time of these packets, granted that the system verify certain criterion. For this purpose, we can see that inherently, the system in consideration is characterized by the following:

1) The packets arrivals of stream 1 are Poisson
2) All packets that enter the system of stream 1, stay in the system until they are served (in other words, there is no blocking, balking or reneging)
3) Packets of stream 1 enter the system and leave one at a time in the order of their arrivals
4) The arrival of packets of stream 1 after time $t$ and the time in the system of any packet of stream 1 arriving before $t$ are independent (i.e. new arriving packets do not influence the time of the system of previous arriving packets)

These conditions are sufficient for the distributional form of Little Law to hold [15]. By defining the Laplace transform of the sojourn time of packets in the system as $P_T(z)$, we can therefore conclude from the distributional form of Little's law that: $P(x) = P_T(\lambda_1 - \lambda_1 x)$. By doing the appropriate change of variable: $x = \frac{\lambda_1 - z}{\lambda_1}$ and after some algebraic manipulations, we get the following:

$$P_T(z) = \frac{\mu_1 \pi_B(0)(w + s + z)}{z^2 + z(w + s + \mu_1 - \lambda_1) + w\mu_1 - \lambda_1 w - \lambda_1 s} \tag{24}$$

One can see that the roots of the denominator verify:

$$\begin{cases} \alpha_1 + \alpha_2 < 0 & \text{if } w + s + \mu_1 > \lambda_1 \\ \alpha_1 \alpha_2 > 0 & \text{if } \mu_1 > \lambda_1(1 + \frac{s}{w}) \end{cases} \tag{25}$$

In other words, the roots of the denominators are negative if and only if these two conditions are satisfied. We recall that the negativity of the poles implies the stability of the queuing system. One can easily see that the second condition is more restrictive (if condition 2 is verified then condition 1 is verified as well). Therefore, for the system to be stable, it is sufficient and necessary to have $\mu_1 > \lambda_1(1 + \frac{s}{w})$ which was previously concluded in Theorem 1. Furthermore, the Laplace transform of sojourn time can be decomposed as follows:

$$P_T(z) = \frac{C_1}{z - \alpha_1} + \frac{C_2}{z - \alpha_2} \tag{26}$$

where $\alpha_1, \alpha_2$ are the roots of the second polynomial in the denominator and $C_1$ and $C_2$ are given by:

$$C_1 = \frac{(\mu_1 - \lambda_1 - \lambda_1 \frac{s}{w})(w + s + \alpha_1)}{(1 + \frac{s}{w})(\alpha_1 - \alpha_2)} \quad (27)$$

$$C_2 = \frac{(\mu_1 - \lambda_1 - \lambda_1 \frac{s}{w})(w + s + \alpha_2)}{(1 + \frac{s}{w})(\alpha_2 - \alpha_1)} \quad (28)$$

By employing the inverse Laplace transform, the proof can be concluded.

## APPENDIX C
## PROOF OF COROLLARY 1

For this theorem, we leverage a characteristic of the proposed Markov chain. First, we define the notion of lumpability.

**Definition 1.** *Suppose that the state-space of a Markov chain is divided into $K$ disjoint subsets of states $\mathcal{P} = \{S_1, \ldots, S_k\}$. A continuous-time Markov chain is said to be lumpable with respect to the partition $\mathcal{P}$ if and only if, for any subset $S_i$ and $S_j$ in $\mathcal{P}$, for any states $c, c' \in S_i$, we have:*

$$\sum_{m \in S_j} q(c, m) = \sum_{m \in S_j} q(c', m) \quad (29)$$

*where $q(c, m)$ is the transition rate between states $c$ and $m$.*

Our proposed chain is characterized by being lumpable with respect to the following partition: $\mathcal{P} = \{S_1, S_2\}$ where $S_1$ is the collection of states where the server is on vacation while $S_2$ contains the states where the server is not on vacation. In fact, as seen in Fig. 7, the total transition rate from any state in the upper part of the chain $S_1$ to the lower one $S_2$ is constant and is equal to $w$. In the other direction, this total transition is equal to $s$. This means that we can reduce our state space to two states: $V$ and $B$ to capture the state of the server. By doing so, we can easily conclude the probability to be in a vacation (and therefore the amount of time the server is on vacation on average) by simply solving for $\pi_V$ in the following equations which concludes our proof:

$$\begin{cases} w\pi_V = s\pi_B \\ \pi_V + \pi_B = 1 \end{cases} \quad (30)$$

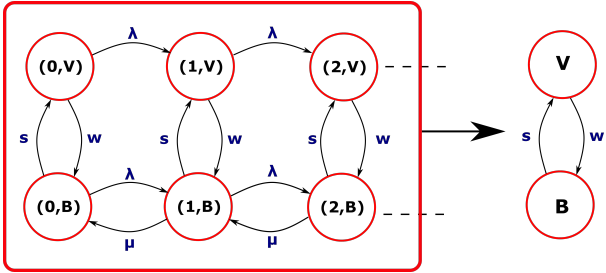

Fig. 7: Lumpable Markov chain

## APPENDIX D
## PROOF OF PROPOSITION 1

To proceed with the proof, we first recall that the sojourn time of packet $j$ denoted as $T_j$ can be decomposed as follows:

$$T_j = B_j + Y_j \quad (31)$$

where $B_j$ is the waiting time of the packet $j$ in the queue and $Y_j$ being the virtual service time of each packet. The virtual service time of packet $j$ is defined as the time elapsed between the time of delivery of packet $j$ and the maximum between the time of delivery of the previous packet $j-1$ or the arrival time of packet $j$[3]:

$$Y_j = D_j - \max(D_{j-1}, t_j) \quad (32)$$

The virtual service time may be different than the actual physical service time due to the possibility of the service of the packet to be interrupted by the vacation. In this case, the calculation of $\mathbb{E}(AT)$ can be decomposed into calculating $\mathbb{E}(AB)$ and $\mathbb{E}(AT)$.

1) $\mathbb{E}(AB)$: The waiting time of a packet can be expressed as $B = (T - A)^+$ where $(.)^+ = \max(., 0)$ [1]. By the total law of expectation, we have the following:

$$\mathbb{E}(AB) = \mathbb{E}(\mathbb{E}(AB|A=a)) =$$
$$\int_0^{+\infty} \int_a^{+\infty} a(t-a) f_T(t) \lambda_1 exp(-\lambda_1 a) dt da \quad (33)$$

By using the results of Theorem 2, we get after some calculations:

$$\mathbb{E}(AB) = \frac{\lambda_1 C_1}{\alpha_1^2 (\alpha_1 - \lambda_1)^2} + \frac{\lambda_1 C_2}{\alpha_2^2 (\alpha_2 - \lambda_1)^2} \quad (34)$$

2) $\mathbb{E}(AY)$: To be able to characterize this term, it is important to discuss the virtual service time of each packet. For this purpose, we define the following event:

$$\Upsilon = \{\text{Arriving packet of stream 1 finds the system}$$
$$\text{in state } \pi_V(0)\}$$

When the complement event $\overline{\Upsilon}$ takes place, one of the following happen:
- The arriving packet finds the server ready to serve it and the physical service of the packet starts
- It finds buffered packets of stream 1 ahead of it

In the first case, the physical service time starts immediately. In the second case, the packet waits for the packets ahead to be served (which is counted in its waiting time $B$). When the packet ahead of it finishes being served, the physical service time of the packet of interest immediately starts. On the other hand, if the event $\Upsilon$ takes place, the arriving packet finds itself at the head of the buffer but the server on vacation and it has to first wait for the vacation to finish before its physical service time starts. We can see from this discussion that the term $\mathbb{E}(AY)$ is tricky to calculate since the arrival time clearly affect the service time of the packet. However, these random variables $A$ and $Y$ are conditionally independent given $\Upsilon$ or $\overline{\Upsilon}$. In fact, the value of $A$ does not affect $Y$ when we know the state of the system the packet arrives to. Based on that, and by leveraging the total law of expectation, we can write the following:

$$\mathbb{E}(AY) = \mathbb{E}(AY|\overline{\Upsilon})P(\overline{\Upsilon}) + \mathbb{E}(AY|\Upsilon)P(\Upsilon)$$
$$= \mathbb{E}(A|\overline{\Upsilon})\mathbb{E}(Y|\overline{\Upsilon})P(\overline{\Upsilon}) + \mathbb{E}(A|\Upsilon)\mathbb{E}(Y|\Upsilon)P(\Upsilon) \quad (35)$$

---
[3]It is worth mentioning that we suppose that the system is in steady state, i.e., the sojourn times and virtual service times of the packets are stochastically identical $T_j \stackrel{st}{=} T_{j-1} \stackrel{st}{=} T$, $B_j \stackrel{st}{=} B_{j-1} \stackrel{st}{=} B$ and $Y_j \stackrel{st}{=} Y_{j-1} \stackrel{st}{=} Y$

In order to deal with these terms, we first recall the PASTA property: an arrival from a Poisson process observes the system as if it was arriving at a random instant of time. Based on that property and by taking into account that the arrival time is independent of the state of the system, we can conclude:

$$\mathbb{E}(AY) = \frac{1}{\lambda_1}\mathbb{E}(Y|\overline{\Upsilon})(1-\pi_V(0)) + \frac{1}{\lambda_1}\mathbb{E}(Y|\Upsilon)\pi_V(0) \quad (36)$$

What remains is to calculate the two terms: $\mathbb{E}(Y|\overline{\Upsilon})$ and $\mathbb{E}(Y|\Upsilon)$. To proceed with that, we elaborate more on what happens when a packet arrives to the system. Suppose that the event $\overline{\Upsilon}$ takes place and the packet is now at head of the buffer: due to the memoryless property of the vacation start timer, two clocks start ticking: an exponential service time $G$ of mean $1/\mu_1$ and an exponential vacation start time $S$ of mean $1/s$. The packet physical service time finishes before the vacation starts with a probability $p = P(G < S) = \frac{\mu_1}{\mu_1+s}$. With a probability $1-p = \frac{s}{\mu_1+s}$, the vacation interrupts the physical service time of the packet for a duration $W$ exponentially distributed with mean $1/w$. After the vacation finishes, the physical service of the packet is resumed. Due to the memoryless property, when the service of the packet is resumed, it is as if a new an exponential service time $G$ of mean $1/\mu_1$ was taken. And therefore, the same probability $1-p$ holds to go for another vacation before the service of the packet finishes. One can see from this that the virtual service time of the packet can include a random number of exponentially distributed vacation times. We therefore define $L$ as the number of vacations that take place during the packet service time. Based on our analysis, $L$ is geometrically distributed with parameter $p$:

$$P(L=i) = (1-p)^i p \quad (37)$$

Knowing that each vacation adds an expected time of $1/w$ to the virtual service time of the packet, and by using the total law of expectation, we have the following:

$$\mathbb{E}(Y|\overline{\Upsilon}) = \sum_{i=0}^{+\infty}\mathbb{E}(Y|\overline{\Upsilon},L=i)P(L=i) = \sum_{i=0}^{+\infty}p(1-p)^i(\frac{1}{\mu_1}+i\frac{1}{w})$$
$$= \frac{1}{\mu_1} + \frac{p}{w}\frac{(1-p)}{p^2} = \frac{1}{\mu_1} + \frac{s}{w\mu_1} = \frac{s+w}{w\mu_1} \quad (38)$$

As for $\mathbb{E}(Y|\Upsilon)$, we know that the only difference between the two virtual service time is that the packet has to wait first for the vacation to finish when the event $\Upsilon$ takes place.

In other words, $\mathbb{E}(Y|\Upsilon) = \mathbb{E}(Y|\overline{\Upsilon}) + \frac{1}{w} = \frac{s+w+\mu_1}{w\mu_1}$. By taking into account that $\pi_V(0) = \frac{s(\mu_1-\lambda_1-\lambda_1\frac{s}{w})}{(\lambda_1+w)(\mu_1(1+\frac{s}{w}))}$, we have the following:

$$\mathbb{E}(AY) = \frac{(s+w+\mu_1)s(\mu_1-\lambda_1-\lambda_1\frac{s}{w})}{(\lambda_1 w\mu_1)(\lambda_1+w)(\mu_1(1+\frac{s}{w}))}$$
$$+ \frac{(s+w)}{\lambda_1 w\mu_1}(1 - \frac{s(\mu_1-\lambda_1-\lambda_1\frac{s}{w})}{(\lambda_1+w)(\mu_1(1+\frac{s}{w}))}) \quad (39)$$

Combining the two results and by applying eq. (3), we end up with the closed form of average age.

## APPENDIX E
## PROOF OF PROPOSITION 2

To proceed with the proof, we study the convexity of the average age expression in eq. (11). The average age $\Delta_1(\lambda_1)$ is decomposed into three parts:
1) $\frac{1}{\lambda_1}$ which is a convex function for any $0 < \lambda_1 < \frac{\mu_1}{1+\frac{s}{w}}$
2) $\frac{\mu_1 s(\mu_1-\lambda_1-\lambda_1\frac{s}{w})+(s+w)(\lambda_1+w)(\mu_1(1+\frac{s}{w}))}{(w\mu_1)(\lambda_1+w)(\mu_1(1+\frac{s}{w}))}$ which can be rewritten as $\frac{a\lambda_1+b}{c\lambda_1+d}$ where:

$$a = \frac{(s+w)^2\mu_1}{w} - \mu_1 s - \mu_1\frac{s^2}{w} \quad b = \mu_1^2 s + (s+w)^2\mu_1$$
$$c = \mu_1^2(s+w) \quad d = w\mu_1^2(s+w) \quad (40)$$

By deriving twice with respect to $\lambda_1$, one can verify that a sufficient and necessary condition for convexity for $\lambda_1 > 0$ is: $bc - ad > 0$. By a simple calculation, this condition can be found to be always verified for any $0 < \lambda_1 < \frac{\mu_1}{1+\frac{s}{w}}$.
3) The first two terms are also two convex functions. The exact calculations are omitted for the sake of space due to the complexity of the objective function, but by deriving twice with respect to $\lambda_1$, the convexity can be verified.
The objective function is therefore a sum of several convex functions and is therefore convex itself. Knowing that $\Delta_1 \longrightarrow +\infty$ when $\lambda_1 \longrightarrow 0$ and when $\lambda_1 \longrightarrow \frac{\mu_1}{1+\frac{s}{w}}$, and by taking into account the convexity of the objective function, we can assert that there exist a *unique* stationary point $0 < \lambda_1^* < \frac{\mu_1}{1+\frac{s}{w}}$ such that $\frac{d\Delta_1(\lambda_1)}{d\lambda_1}|_{\lambda_1=\lambda_1^*} = 0$. This observation renders the Lagrangian approach to the problem unnecessary as this stationary point is the *unique* minimizer of the objective function in the interval of interest.